\documentclass[structabstract]{aa}  
\usepackage{longtable,lscape,graphicx,float,amsmath,amssymb,mathrsfs,txfonts}
\usepackage{multirow,natbib} 
\usepackage{txfonts}
\authorrunning{J.-B. Bossa et al.}
\titlerunning{Porosity measurements of interstellar ice mixtures using optical laser interference and EMAs}
\begin{document}
   \title{Porosity measurements of interstellar ice mixtures using optical laser interference and extended effective medium approximations}

   \subtitle{}

  \author{J.-B. Bossa \inst{1}, K. Isokoski \inst{1}, D. M. Paardekooper \inst{1}, M. Bonnin \inst{1}, E. P. van der Linden \inst{1}, T. Triemstra \inst{1}, S. Cazaux \inst{2},\\ A. G. G. M. Tielens \inst{3} \and H. Linnartz \inst{1}}

   \institute{Sackler Laboratory for Astrophysics, Leiden Observatory, Leiden University, P.O. Box 9513, NL 2300 RA Leiden, The Netherlands.\\
              \email{bossa@strw.leidenuniv.nl; linnartz@strw.leidenuniv.nl }
         \and
             Kapteyn Astronomical Institute, P.O. Box 800, NL 9700 AV Groningen, The Netherlands.\\
         \and 
             Leiden Observatory, Leiden University, P.O. Box 9513, NL 2300 RA Leiden, The Netherlands.\\  
             }



   \date{Received 27 August 2013 / Accepted 29 November 2013}

  
  \abstract
   {}
   {This article aims to provide an alternative method of measuring the porosity of multi-phase composite ices from their refractive indices and of characterising how the abundance of a premixed contaminant (e.g., CO$_{2}$) affects the porosity of water-rich ice mixtures during omni-directional deposition.}
   {We combine optical laser interference and extended effective medium approximations (EMAs) to measure the porosity of three astrophysically relevant ice mixtures: H$_{2}$O:CO$_{2}$=10:1, 4:1, and 2:1. Infrared spectroscopy is used as a benchmarking test of this new laboratory-based method.}
   {By independently monitoring the O-H dangling modes of the different water-rich ice mixtures, we confirm the porosities predicted by the extended EMAs. We also demonstrate that CO$_2$ premixed with water in the gas phase does not significantly affect the ice morphology during omni-directional deposition, as long as the physical conditions favourable to segregation are not reached. We propose a mechanism in which CO$_2$ molecules diffuse on the surface of the growing ice sample prior to being incorporated into the bulk and then fill the pores partly or completely, depending on the relative abundance and the growth temperature.}
  {}


   \maketitle
%

\section{Introduction}
\label{section1}

Amorphous solid water and carbon dioxide, two of the most abundant ices, are thought to form simultaneously on interstellar grain surfaces present in quiescent clouds and star forming regions \citep{Gerakines96,Palumbo98,Jamieson06,Ioppolo09,Noble11}. CO$_{2}$ ice is ubiquitous and the abundances range from $\sim$7 to $\sim$38 \% relative to H$_{2}$O ice, depending on the targeted source type \citep{Gibb04,Oberg2011}. Numerous laboratory experiments have added to our present understanding of the infrared signatures and the thermal history of H$_{2}$O- and CO$_{2}$-bearing ice mantles \citep{Ehrenfreund99,Palumbo06,Hodyss08,Mate08,Oberg09a,Fayolle11a}.\\
\indent The porosity of pure H$_{2}$O ice has been extensively studied in the laboratory \citep{Stevenson99, Kimmel01,Dohnalek03}, but quantitative information on the porosity of water-rich \textit{ice mixtures} is lacking. Porous ices are expected to increase the efficiency of solid state astrochemical processes since they are chemically more reactive than compact (non-porous) ices; they provide larger effective surface areas for catalysis, for the freeze-out of additional atoms and molecules, and for the trapping of volatiles. Understanding how an impurity influences the H$_{2}$O porosity during ice growth is therefore crucial for predicting the chemical evolution of interstellar ice analogues in different astronomical environments.\\
\indent The degree of porosity of inter- and circumstellar ices is still an open question. Remote observations of interstellar ices \citep{Keane01} and laboratory studies of the stability of porous H$_{2}$O ice samples converge to the same conclusion that porosity is rare in space owing to external influences, such as thermal annealing, ion impact, and VUV irradiation, or to H-atom bombardment \citep{Palumbo06,Raut07,Raut08,Palumbo10,Accolla11}. The missing O-H dangling features in astronomical spectra have been taken as a proof for compact amorphous solid water. However, care is needed, since laboratory data show that the absence of the O-H dangling modes does not necessarily imply the full absence of porosity \citep{Raut07}. Moreover, different ways of energetic processing in space can desorb icy material and redistribute them on remaining cold surfaces, thus providing several alternative routes to different ice porosities. These processes include photodesorption \citep{Greenberg73,Oberg09b,Oberg09c,Fayolle11b}, thermal cycling of material between the diffuse interstellar medium and dense clouds \citep{McKee89}, vertical and radial transport of ices within disks \citep{Williams2011}, exothermic solid state reactions \citep{Cazaux2010,Dulieu2013}, and sputtering of material in shock regions \citep{Bergin1999}.\\  
\indent In the laboratory, vapour deposition of one gas phase constituent (e.g., H$_{2}$O) on a cold substrate results in a two-phase composite ice sample by taking the presence of pores into account, which are inevitably formed during growth \citep{Brown96,Dohnalek03,Mate12}. The resulting porosity depends on experimental conditions, such as the growth temperature, the deposition rate, and the growth angle. In the same way, a three-phase composite ice sample is expected when depositing two gas phase constituents (e.g., H$_{2}$O and CO$_{2}$) on a cold substrate. The resulting porosity may also depend on new parameters, such as the abundance and the nature of the premixed contaminant.\\
\indent The goal of the present study is to characterise how the abundance of CO$_{2}$ affects the porosity of thick ($>$100 ML) H$_{2}$O:CO$_{2}$ ice samples, grown by background deposition at different growth temperatures. For that purpose we combine optical laser interference with two distinct effective medium approximations (EMAs), namely Maxwell-Garnett and Bruggeman. EMAs have already proven to be fairly good optical constant predictions in the mid infrared for dirty ices \citep{Mukai84,Mukai86,Preibisch1993}. Here for the first time, these well known EMAs are used to characterise the porosity of multi-phase composite ices. Section \ref{section2} describes details on the experimental methods and data interpretation.  Section \ref{section3} presents both laboratory results and model predictions on the porosity of pure H$_2$O, and H$_{2}$O:CO$_{2}$=10:1, 4:1, and 2:1 ice samples. Finally, the results are discussed in section \ref{section4}, that includes infrared spectroscopy as a benchmarking test of the porosity predictions. Here also the astronomical relevance of this work is discussed. A summary and concluding remarks are given in the final section.




\section{Experimental methods}
\label{section2}
\subsection{Background deposition and high-vacuum set-up}
The experimental set-up and procedures for monitoring the ice thickness during deposition have been described previously \citep{Bouwman07,Bossa12}. In brief, different ice samples (pure H$_2$O,  H$_{2}$O:CO$_{2}$=10:1, 4:1, and 2:1) are grown by background deposition on a cryogenically cooled silicon substrate located in the center of a high-vacuum chamber (2$\times$10$^{-7}$ Torr at room temperature). A gas inlet tube is directed away from the substrate, which allows the gas phase molecules to impinge the surface with random trajectories, thus providing porous structures and ensuring an uniform ice growth. A large volume (2 L) gas reservoir together with an aperture adjusted leak valve are used to ensure a constant deposition rate. The silicon substrate is mounted on the tip of a closed-cycle helium cryostat that, in conjunction with resistive heating, allows an accurate temperature control from 18 to 300 K with a precision of 0.1 K. The gases that we use include carbon dioxide (CO$_2$) (Praxair, purity 99.998 \%) and milli-Q grade water (H$_2$O) that is purified by several freeze-thaw cycles prior to deposition. The relative molecular abundances in the gas phase are obtained by a standard manometric technique with an absolute precision of 10 \%. The final mixture ratio accuracy in the solid phase is estimated of the order of 30 \% based on infrared spectroscopy and the integrated absorption coefficients from the literature \citep{Schutte95,Oberg07}.
\subsection{Optical laser interference}
\indent The optical interference experiments are performed using an intensity stabilised red ($\lambda$\,=\,632.8 nm) helium-neon (He-Ne) laser (Thorlabs HRS015). The laser beam is s-polarized (perpendicular) with respect to the plane of incidence, and strikes the substrate surface at an incident angle $\theta_{0}$\,$\simeq$\,45\,$^\circ$. The reflected light is thereafter collected and converted to an analogue signal by an amplified photodiode (Thorlabs PDA36A). The photodiode signal and the substrate temperature are recorded simultaneously as a function of time using LabVIEW 8.6 (National Instruments) at a 0.5 Hz sampling rate. Optical interference versus deposition time results in a signal intensity modulation due to constructive and destructive interferences \citep{Brown96, Westley98, Dohnalek03}. Ice sample depositions are typically stopped when the signal is located at an upward slope, e.g., half-way between the third destructive interference and the third constructive interference \citep{Bossa12}. The resulting ice thicknesses are below 1 $\mu$m and the typical deposition time is ranging between $\sim$35 and $\sim$55 minutes, depending on the mixture and the growth temperature.\\

\begin{figure}[]
\centering
\includegraphics[width=8.5cm]{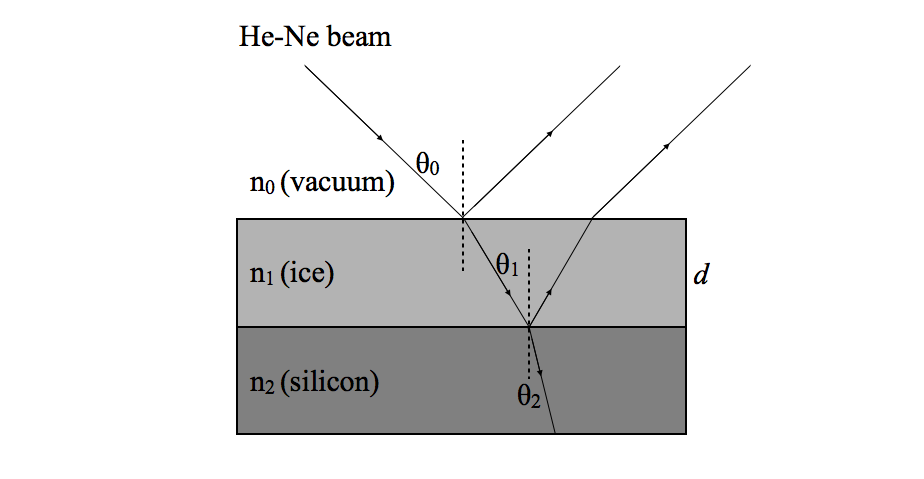} \caption{Schematic of the three-phase system layered structure.}
\label{fig7}
\end{figure}

\indent We first use a three-phase layered model (vacuum, ice, and silicon as depicted in Fig. \ref{fig7}) to derive the refractive indices of the ice samples, $n_{1}$(T), that depend on the growth temperature. In this way, we approximate that ice samples are homogeneous, i.e, the He-Ne light is only reflected off the two vacuum/ice and ice/silicon interfaces only. The total reflection coefficient $R[d,n_{1}(\textrm{T})]$ can be written as a function of the Fresnel reflection coefficients according to the relation \citep{Westley98,Dohnalek03}    
\begin{equation}
\label{eq15}
R[d,n_{1}(T)] = \frac{r_{01}(\textrm{T}) + r_{12}(\textrm{T})\,e{^{-i2\beta(\textrm{T})}}}{1 + r_{01}(\textrm{T})\,r_{12}(\textrm{T})\,e{^{-i2\beta(\textrm{T})}}}.
\end{equation}
\noindent The exponential term $\beta(\textrm{T})$ describes the phase change of the light as it passes through the ice sample of thickness $d$
\begin{equation}
\label{eq16}
\beta(\textrm{T}) = \frac{2\pi d}{\lambda}\,n_{1}(\textrm{T})\, \cos \theta_{1}.
\end{equation}
 \noindent The Fresnel reflection coefficients $r_{01}$(\textrm{T}) and $r_{12}$(\textrm{T}) are associated with the vacuum/ice and ice/silicon interfaces, respectively. They are also a function of the complex refractive indices  n$_{0}$ (vacuum), $n_{1}$(T) (ice), and n$_{2}$ (silicon). For s-polarized light, the Fresnel reflection coefficients are
\begin{equation}
\label{eq17}
r_{01s}(\textrm{T}) = \frac{ n_{0}\,\cos{\theta_{0}} - n_{1}(\textrm{T})\,\cos{\theta_{1}} }{ n_{0}\,\cos{\theta_{0}} + n_{1}(\textrm{T})\,\cos{\theta_{1}} },
\end{equation}
\begin{equation}
\label{eq18}
r_{12s}(\textrm{T}) = \frac{ n_{1}(\textrm{T})\,\cos{\theta_{1}} - n_{2}\,\cos{\theta_{2}} }{ n_{1}(\textrm{T})\,\cos{\theta_{1}} + n_{2}\,\cos{\theta_{2}} }.
\end{equation}
\noindent The incident angles ($\theta_{0}$, $\theta_{1}$, and $\theta_{2}$) and the complex refractive indices ($n_{0}$, $n_{1}$(T), and $n_{2}$) are related through Snell's law. We use constant refractive indices n$_{0}$\,=\,1 (vacuum), and n$_{2}$\,=\,$3.85\,-\,0.07i$ (silicon) \citep{Mottier81}. The values of $n_{1}$(T) (ice) are discussed in the next section. An independent control experiment is performed (see subsection \ref{section23}) that allows a qualitative comparison with the results obtained from interpreting the interference data.

\subsection{Infrared spectroscopy}
\label{section23}
\indent Infrared spectroscopy is used after each background deposition in order to compare the amount of pores present in the different ice samples. Optical laser interference and infrared spectroscopy cannot be performed simultaneously because of geometrical restrictions in the HV set-up \citep{Bossa12}. Therefore separate experiments are performed but with identical deposition procedures. We focus on the 3750 -- 3550 cm$^{-1}$ range that covers the combination/overtone modes of carbon dioxide and the O-H dangling modes of water. The outcomes are then used as a benchmarking test of the two extended EMAs. Infrared spectra are obtained with a Fourier Transform Infrared spectrometer (Varian 670-IR FTIR) and recorded in transmission mode between 4000 and 800 cm$^{-1}$. The infrared beam transmits through the ice sample and the silicon substrate at an incident angle $\simeq$\,45\,$^\circ$. An infrared spectrum has a 1 cm$^{-1}$ resolution and is averaged over 256 interferograms. The FTIR is flushed with dry air to minimise background fluctuations due to atmospheric absorptions. Background spectra are acquired at the specific growth temperature prior to deposition for each experiment.

\section{Results}
\label{section3}
\subsection{Temperature dependent refractive indices of porous H$_2$O and porous H$_{2}$O:CO$_{2}$ ice samples}
\label{section6}
\begin{figure}[]
\centering
\includegraphics[width=8.5cm]{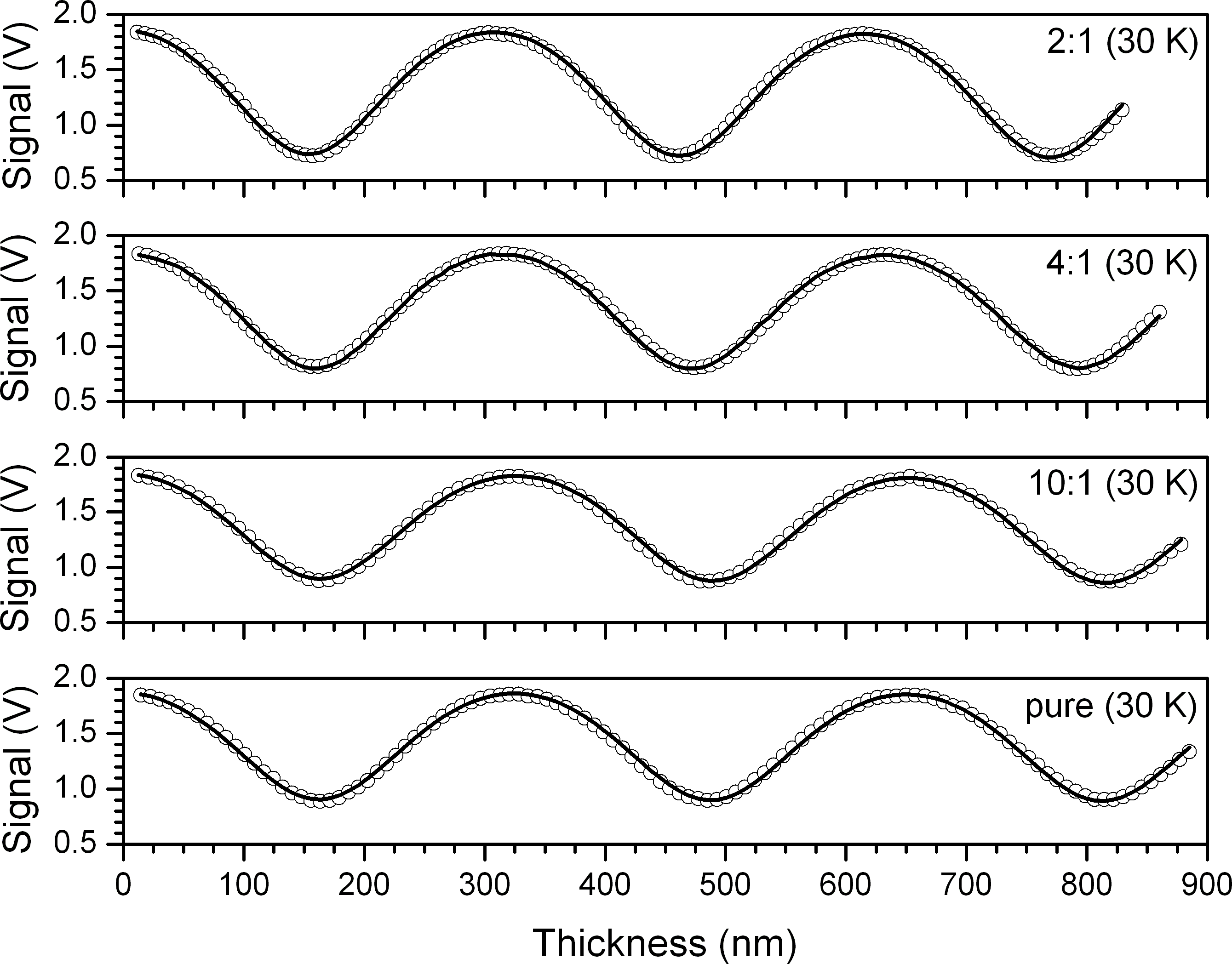} \caption{Optical interference fringe pattern obtained during the background deposition of porous H$_2$O (pure), porous H$_{2}$O:CO$_{2}$=10:1, 4:1, and 2:1 ice samples at 30 K. The open circles correspond to the reduced experimental data points and the solid lines represent the fits.}
\label{fig4}
\end{figure}

Measurements of the refractive indices of porous H$_2$O, and porous H$_{2}$O:CO$_{2}$=10:1, 4:1, and 2:1 ice samples are performed in the 30 -- 70 K growth temperature range. Beyond 70 K, carbon dioxide molecules barely stick on the silicon substrate, making the determination of the final H$_{2}$O:CO$_{2}$ ratio difficult. The refractive indices $n_{1}$(T) result from fitting the complete interference fringe pattern to the reflectance signal $\left| R[d,n_{1}(T)]  \right|^{2}$. The fitting procedure is driven by Matlab 7.9.0 (R2009b), and uses the Nelder-Mead optimisation algorithm \citep{Lagarias98}. The fitting parameters are a complex refractive index, a linear deposition rate, and a scale factor that translates light intensity to Volts. Examples of interference fringe patterns obtained during background deposition (reduced data points, open circles) and corresponding fits (solid lines) are shown in Fig. \ref{fig4} for the four ice samples grown at 30 K. In general, the interference data exhibit a small amplitude damping with ongoing deposition, most likely caused by surface roughness and/or cracks \citep{Baragiola03}. The distance between subsequent minima and maxima is nearly constant, indicating that (i) the density does not change significantly during deposition \citep{Westley98}, and (ii) the deposition rate is constant. The values of $n_{1}$(T) obtained from fitting the reflectance are given in Table \ref{tab1}. The imaginary components of the complex refractive indices are approximately zero, indicating that the He-Ne light absorption in the different ice samples is negligible. \\

\indent The refractive indices of porous H$_2$O ice samples deposited below 70 K are in agreement with previously published values \citep{Dohnalek03}, hence confirming our fitting procedure. From Table \ref{tab1}, we observe that the refractive indices increase nearly linearly with increasing growth temperature, which is consistent with the hypothesis that an increase in the refractive index is mainly due to a decrease of the porosity \citep{Brown96,Dohnalek03,Mate12}. The refractive indices of H$_{2}$O:CO$_{2}$=10:1 ice samples are comparable to the ones of porous H$_2$O ice samples, whereas the real components of the refractive indices increase with higher CO$_{2}$ abundances. A maximum of $n_{1}$(T) $\simeq$ 1.3 is reached for H$_{2}$O:CO$_{2}$=2:1 ice samples grown between 50 and 60 K. These changes in the refractive indices with growth temperatures and CO$_{2}$ abundances are likely due to changes in the dielectric properties of the ices. In the three-phase layered model, all the involved interfaces need to be taken into account and because we first approximate our ice sample as being homogeneous, the bulk water ice/pores, bulk water ice/CO$_{2}$, and pores/CO$_{2}$ interfaces are neglected. In order to take this effect into account, the next sections treat the different ice samples as heterogeneous materials.

\begin{table*}[ht!]
\caption{Refractive indices $n_{1}$(T) obtained from fitting the complete interference fringe pattern to the reflectance signal for different ice samples deposited at different growth temperatures. The error in the refractive indices is given at 3$\sigma$ confidence level.}  
\label{tab1}      
\centering                                      
\begin{tabular}{c l l l l}          
\hline\hline                       
Temperature     & H$_2$O & H$_{2}$O:CO$_{2}$ & H$_{2}$O:CO$_{2}$ &  H$_{2}$O:CO$_{2}$  \\    
 (K)                   &  pure       & 10:1                           & 4:1                            & 2:1\\ 
\hline                                   
    30 & 1.203 $\pm$ 0.004& 1.201 $\pm$ 0.004  & 1.224 $\pm$ 0.004  & 1.248 $\pm$ 0.004 \\      
    40 & 1.211 $\pm$ 0.005& 1.208 $\pm$ 0.003  & 1.234 $\pm$ 0.004  & 1.274 $\pm$ 0.004 \\
    50 & 1.214 $\pm$ 0.005& 1.213 $\pm$ 0.004  & 1.236 $\pm$ 0.004  &  1.299 $\pm$ 0.005\\
    60 & 1.216 $\pm$ 0.005& 1.215 $\pm$ 0.004  & 1.242 $\pm$ 0.005  &  1.300 $\pm$ 0.006 \\
    70 & 1.219 $\pm$ 0.005& 1.223 $\pm$ 0.005  & 1.250 $\pm$ 0.005  &   1.291 $\pm$ 0.006\\
\hline                                             
\end{tabular}
\end{table*}

\subsection{Porous H$_2$O ice samples as heterogeneous materials}
In this section we visualise a porous H$_2$O ice sample as a heterogeneous material, i.e, two different dielectric materials compose the ice sample: water and pores. The refractive indices of an ice mixture can be predicted from the refractive indices of its constituents as long as there are no interactions (physical and/or chemical) between the constituents \citep{Mukai86}. For that, different EMAs have been developed many years ago, e.g., Maxwell-Garnett (1904) and Bruggeman (1935). The following shows the predictions of these two distinct EMAs for porous H$_2$O ice samples deposited at different growth temperatures. We then compare the predictions with the experimental data obtained in section \ref{section6}. 

\subsubsection{The Maxwell-Garnett EMA}

\begin{figure}[]
\centering
\includegraphics[width=8.5cm]{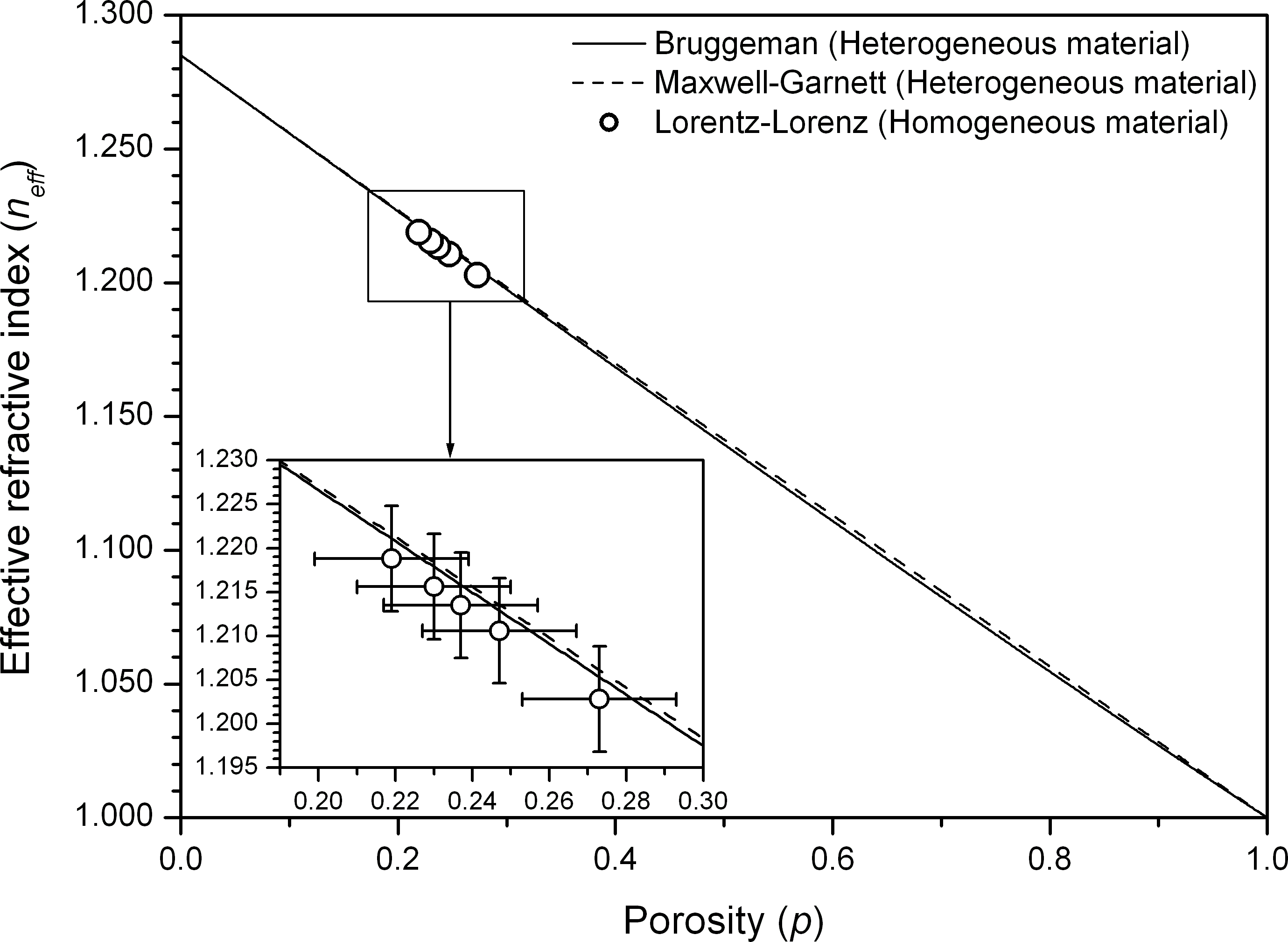} \caption{Effective refractive index ($n_{eff}$) predicted by the Bruggeman EMA (solid line), and by the Maxwell-Garnett EMA (dash dot) as a function of porosity. The open circles indicate the refractive index and the corresponding porosity obtained by laser optical interference and by the Lorentz-Lorenz equation (Eq. \ref{eq1}), respectively. The inset presents a zoom-in of the low porosity range.} 
\label{fig1}
\end{figure}

The Maxwell Garnett EMA treats the system asymmetrically: one can visualise the system as inclusions ($incl$) evenly distributed in a host medium ($h$). The inclusions are assumed to be spheres (or ellipsoids) of a size and a separation distance smaller than the optical wavelength. Under these conditions, one can treat a porous H$_2$O ice sample as an effective medium, characterised by an effective dielectric constant ($\epsilon_{eff}$) that satisfies the equation \citep{MG1904, MG1906}
\begin{equation}
\label{eq8}
\frac{\epsilon_{eff} - \epsilon_{h}}{\epsilon_{eff} + 2\epsilon_{h}} = f_{incl} \times \frac{\epsilon_{incl} - \epsilon_{h}}{\epsilon_{incl} + 2\epsilon_{h}}, 
\end{equation}
with $\epsilon_{incl}$ and $\epsilon_{h}$ the dielectric constants of the inclusions and the host medium, respectively, and $f_{incl}$ the volume fraction of the inclusions. We need now to define the spherical inclusions and the host material. Since the average density of vapour deposited ice is relatively close to the intrinsic density of bulk water ice \cite{Raut07}, we assume that pores are less abundant than the bulk water ice, thus we arbitrarily define the spheres of air (pores) as the inclusions, and the bulk water ice as the host material. Hence for a porous H$_2$O ice sample of porosity $p$, deposited at a growth temperature T, the effective dielectric constant $\epsilon_{eff}$(T) can be determined by solving the following equation for different porosities ($0 \le p \le 1$)
\begin{equation}
\label{eq6}
\frac{\epsilon_{eff}(\textrm{T}) -  \epsilon_{bulk\,water}}{\epsilon_{eff}(\textrm{T}) + 2\epsilon_{bulk\,water}} = p \times \frac{\epsilon_{pores}  -  \epsilon_{bulk\,water}}{\epsilon_{pores}  + 2 \epsilon_{bulk\,water}}.
\end{equation}
\noindent We assume that the dielectric constant of the pores equals 1 and that the He-Ne light absorption in the ice is negligible, then Eq. \ref{eq6} becomes 
\begin{equation}
\label{eq7}
\frac{n_{eff}^{2}(\textrm{T}) - n_{i}^{2}}{n_{eff}^{2}(\textrm{T}) + 2n_{i}^{2}} = p \times \frac{1 - n_{i}^{2}}{1 + 2 n_{i}^{2}},
\end{equation}
where $n_{i}$ and $n_{eff}$ correspond to the intrinsic refractive index of the bulk water ice (i.e., not including pores) and the effective refractive index of the heterogeneous material, respectively. The $n_{i}$ value can be found in the literature \citep{Brown96,Westley98, Dohnalek03}, and we use $n_{i}$ = 1.285 from \citep{Dohnalek03}. This value comes from the intrinsic density (i.e., not including pores) of the low-density amorphous form (I$_{a}$l, 0.94 g cm$^{-3}$) \citep{Jenniskens94,Jenniskens95}. The low-density amorphous form is observed for growth temperatures between 30 and 135 K. Therefore, we assume that the variation of the dielectric constant of the bulk water ice is negligible within the 30 -- 70 K growth temperature range. Fig. \ref{fig1} shows the effective refractive indices predicted by the Maxwell-Garnett EMA (dash dot, heterogeneous material). In theory, the porosity can be adjusted continuously from $p$ = 0, ($n_{eff} = n_{i} = 1.285$) to $p$ = 1 ($n_{eff} = n_{0} = 1$). In practice, the porosity never exceeds 0.4 following the background deposition procedure. Predictions from spherical units do not show noticeable differences compared to the ellipsoid geometry model (not shown here).\\

\indent We directly compare the solutions from Eq. \ref{eq7} with the refractive indices, $n_{1}$(T), measured by laser optical interference (using the three-phase layered model) at different growth temperatures (see section \ref{section6}, and Eqs. \ref{eq15} -- \ref{eq18}). From each measured refractive index, one can derive the corresponding porosity, $p$, using the Lorentz-Lorenz equation \citep{Westley98,Dohnalek03,Raut07}  
\begin{equation}
\label{eq1}
p=1 -  \Bigg( \frac{n_{1}^{2}\textrm{(T)}-1}{n_{1}^{2}\textrm{(T)}+2} \times \frac{n_{i}^{2}+2}{n_{i}^{2}-1} \Bigg).
\end{equation}

 \noindent By using the three-phase layered model and the Lorentz-Lorenz equation, the system is regarded as a homogeneous material based on the size limit set by the optical wavelength \citep{Aspnes82,Dohnalek03}. Fig. \ref{fig1} shows the measured refractive indices of porous H$_2$O ice samples as a function of the derived porosities (open circles, homogeneous material), in comparison to the Maxwell-Garnett EMA (dash dot, heterogeneous material). The lowest porosity is obtained with the maximum growth temperature of 70 K, and the predicted values from the Maxwell-Garnett EMA are similar to the predicted values from the Lorentz-Lorenz equation. At this stage, the small but systematic offset between theory and values predicted by the Lorentz-Lorenz equation cannot be explained. A direct experimental measurement of the density using a quartz crystal microbalance may add information to the experiment described here.

\subsubsection{The Bruggeman EMA}

In contrast to the Maxwell-Garnett EMA, the Bruggeman EMA treats the system symmetrically: one can visualise the system as spheres of air (pores) and spheres of bulk water ice embedded in an effective medium, characterised by an effective dielectric constant ($\epsilon_{eff}$) that satisfies the equation \citep{BR1935}  \\
\begin{equation}
\label{eq2}
\displaystyle \sum_{j=1}^{2} f_{j}\,\times\,  \Big(\frac{\epsilon_{j} - \epsilon_{eff}}{\epsilon_{j} + 2\epsilon_{eff}}\Big)= 0,
\end{equation}
with the condition
\begin{equation}
\label{eq3}
\displaystyle \sum_{j=1}^{2} f_{j} = 1,
\end{equation}
where $\epsilon_{j}$ and  $f_{j}$ represent the dielectric constant and volume fraction of constituent $j$, respectively. Hence for a porous H$_2$O ice sample of porosity $p$, deposited at a growth temperature T, the effective dielectric constant $\epsilon_{eff}$(T) can be determined by solving the following equation for different porosities ($0 \le p \le 1$)
\begin{equation}
\label{eq4}
(1-p) \times \Bigg(\frac{\epsilon_{bulk\,water} - \epsilon_{eff}(\textrm{T})}{\epsilon_{bulk\,water} + 2\epsilon_{eff}(\textrm{T})}\Bigg) + p \times \Bigg(\frac{\epsilon_{pores} - \epsilon_{eff}(\textrm{T})}{\epsilon_{pores} + 2\epsilon_{eff}(\textrm{T})}\Bigg)= 0.
\end{equation}
As previously, we assume that the dielectric constant of the pores equals 1 and that the He-Ne light absorption in the ice sample is negligible, then Eq. \ref{eq4} becomes
\begin{equation}
\label{eq5}
(1-p) \times \Bigg(\frac{n_{i}^{2} - n_{eff}^{2}(\textrm{T})}{n_{i}^{2} + 2n_{eff}^{2}(\textrm{T})}\Bigg) + p \times \Bigg(\frac{1 - n_{eff}^{2}(\textrm{T})}{1 + 2n_{eff}^{2}(\textrm{T})}\Bigg)= 0.
\end{equation}

\noindent Fig. \ref{fig1} also shows the effective refractive indices predicted by the Bruggeman EMA (solid line, heterogeneous material), in comparison to the Maxwell-Garnett EMA (dash dot, heterogeneous material) and the Lorentz-Lorenz equation (open circles, homogeneous material). We observe that the predicted values from the Bruggeman EMA are similar to the predicted values from the Maxwell-Garnett EMA.\\
\indent There is no general answer to which is the best EMA to characterise a composite material. A thorough comparison between theory and optical experiments is needed to determine which is the most suitable model. Most heterogeneous materials can be approximated by the two presented EMAs \citep{Niklasson81}. In general, the Maxwell Garnett EMA is expected to be valid with inclusions (e.g., pores) occupying low volume fractions, and the Bruggeman EMA is frequently used to describe both surface roughness and porosity. Therefore, it is not surprising that for low porous ($0 < p < 0.3$) H$_2$O ice samples, Bruggeman and Maxwell-Garnett EMAs give very similar results.\\
\indent To summarise, we observe that the data obtained by the Lorentz-Lorenz equation treating the porous H$_2$O ice samples as homogeneous materials agree well with the predictions based on the Maxwell-Garnett and the Bruggeman EMAs, in which the same ice samples are treated as heterogeneous materials. Therefore, in the case of porous H$_2$O ice samples where the bulk water ice/pores interfaces are present, the three-phase layered model is still valid, and we can assume that $n_{eff}$(T) = $n_{1}$(T). 

\subsection{Porous H$_2$O:CO$_2$ ice samples as heterogeneous materials}
In this section we visualise a porous H$_2$O:CO$_2$ ice sample as a heterogeneous material, i.e, three different dielectric materials compose the ice sample: water, carbon dioxide, and pores. We assume that the constituents do not strongly interact with each other. Previous research focused on extended effective medium approximations with the aim of predicting the optical constants of systems of three components \citep{Wachniewski86, Jayannavar91,Nicorovici95, Luo97}. Extrapolating from porous H$_2$O ice samples discussed above, we assume in this section that the refractive indices measured by laser optical interference correspond to the effective refractive indices. Using this assumption, we are able to predict how the abundance of CO$_2$ affects the porosities of H$_2$O:CO$_2$ ice samples deposited at different growth temperatures. 

\subsubsection{The extended Maxwell Garnett EMA}
\label{extendedMG}
The extended model proposed by \citep{Luo97} is of particular interest in this section since a three-component composite material is visualised as a separated-grain structure in which two different particles ($A$ and $B$) are dispersed in a continuous host of dielectric medium $C$. In the following, the three different components $A$, $B$, $C$ are characterised by their dielectric constants $\epsilon_{A}$, $\epsilon_{B}$, and $\epsilon_{C}$, their refractive indices $n_{A}$, $n_{B}$, and $n_{C}$, and their volume fractions $f_{A}$, $f_{B}$, and $f_{C}$ (with the condition: $f_{A}+f_{B}+f_{C}=1$). We arbitrarily define the spheres of air (pores) as $A$, the spheres of bulk carbon dioxide ice as $B$, and the bulk water ice as the host material $C$. In this way, the volume fraction $f_{A}$ corresponds to the porosity. Hence for a porous H$_2$O:CO$_2$ ice sample of porosity $f_{A}$, deposited at a growth temperature T, the effective dielectric constant $\epsilon_{eff}$(T) can be determined by solving the following equation for different porosities ($0.001 \le f_{A} \le 0.998$)  \citep{Luo97}
\begin{equation}
\begin{split}
\label{eq9}
& p_{A} \frac{(\epsilon_{C}-\epsilon_{eff})(\epsilon_{A}+2\epsilon_{C})+f_{AB}(2\epsilon_{C} + \epsilon_{eff})(\epsilon_{A}-\epsilon_{C})}{(\epsilon_{C}+2\epsilon_{eff})(\epsilon_{A}+2\epsilon_{C})+f_{AB}(2\epsilon_{C} - 2\epsilon_{eff})(\epsilon_{A}-\epsilon_{C})} \\ &+ p_{B} \frac{(\epsilon_{C}-\epsilon_{eff})(\epsilon_{B}+2\epsilon_{C})+ f_{AB}(2\epsilon_{C}+\epsilon_{eff})(\epsilon_{B}-\epsilon_{C})}{(\epsilon_{C}+2\epsilon_{eff})(\epsilon_{B}+2\epsilon_{C})+ f_{AB}(2\epsilon_{C} - 2\epsilon_{eff})(\epsilon_{B}-\epsilon_{C})} \\ &= 0, 
\end{split}
\end{equation}
with $p_{A} = f_{A}/(f_{A} + f_{B})$, $p_{B} = f_{B}/(f_{A} + f_{B})$, and  $f_{AB} = f_{A} + f_{B}$. The mathematical approach and methodology can be found in more detail in \citep{Luo97}, based on the research of \citep{Niklasson81}. We assume again that the dielectric constant of the pores equals 1 and that the He-Ne light absorption in the ice sample is negligible, then Eq. \ref{eq9} becomes
\begin{equation}
\begin{split}
\label{eq10}
& p_{A} \frac{(n_{C}^{2}-n_{eff}^{2})(1+2n_{C}^{2})+f_{AB}(2n_{C}^{2} + n_{eff}^{2})(1-n_{C}^{2})}{(n_{C}^{2}+2n_{eff}^{2})(1+2n_{C}^{2})+f_{AB}(2n_{C}^{2} - 2n_{eff}^{2})(1-n_{C}^{2})} \\ &+ p_{B} \frac{(n_{C}^{2}-n_{eff}^{2})(n_{B}^{2}+2n_{C}^{2})+ f_{AB}(2n_{C}^{2}+n_{eff}^{2})(n_{B}^{2}-n_{C}^{2})}{(n_{C}^{2}+2n_{eff}^{2})(n_{B}^{2}+2n_{C}^{2})+ f_{AB}(2n_{C}^{2} - 2n_{eff}^{2})(n_{B}^{2}-n_{C}^{2})} \\ &= 0, 
\end{split}
\end{equation}
where $n_{B}$ and $n_{C}$ correspond to the intrinsic refractive indices of the bulk carbon dioxide ice (i.e., not including pores), and the bulk water ice ($n_{C}=1.285$), respectively. We take $n_{B}$ from the literature, which is 1.41 at 632.8 nm \citep{Warren86}. Above about 70 K this value agrees well with values given in the literature for the bulk carbon dioxide crystal \citep{Schulze80}. In addition, we assume that the variation of the dielectric constant of the bulk CO$_2$ ice is negligible in the 30 -- 70 K growth temperature range.\\

\begin{figure}[]
\centering
\includegraphics[width=8.5cm]{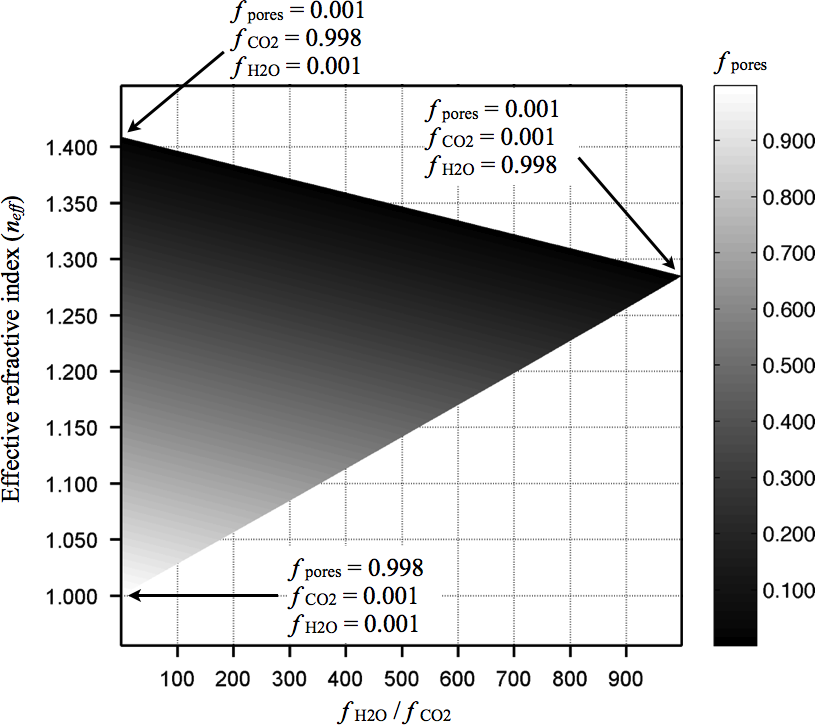} \caption{Effective refractive index ($n_{eff}$) predicted by the extended Maxwell-Garnett EMA as a function of H$_{2}$O:CO$_{2}$ volume fraction ratios and porosity (colorbar).} \label{fig2}
\end{figure}

\begin{figure}[]
\centering
\includegraphics[width=8.5cm]{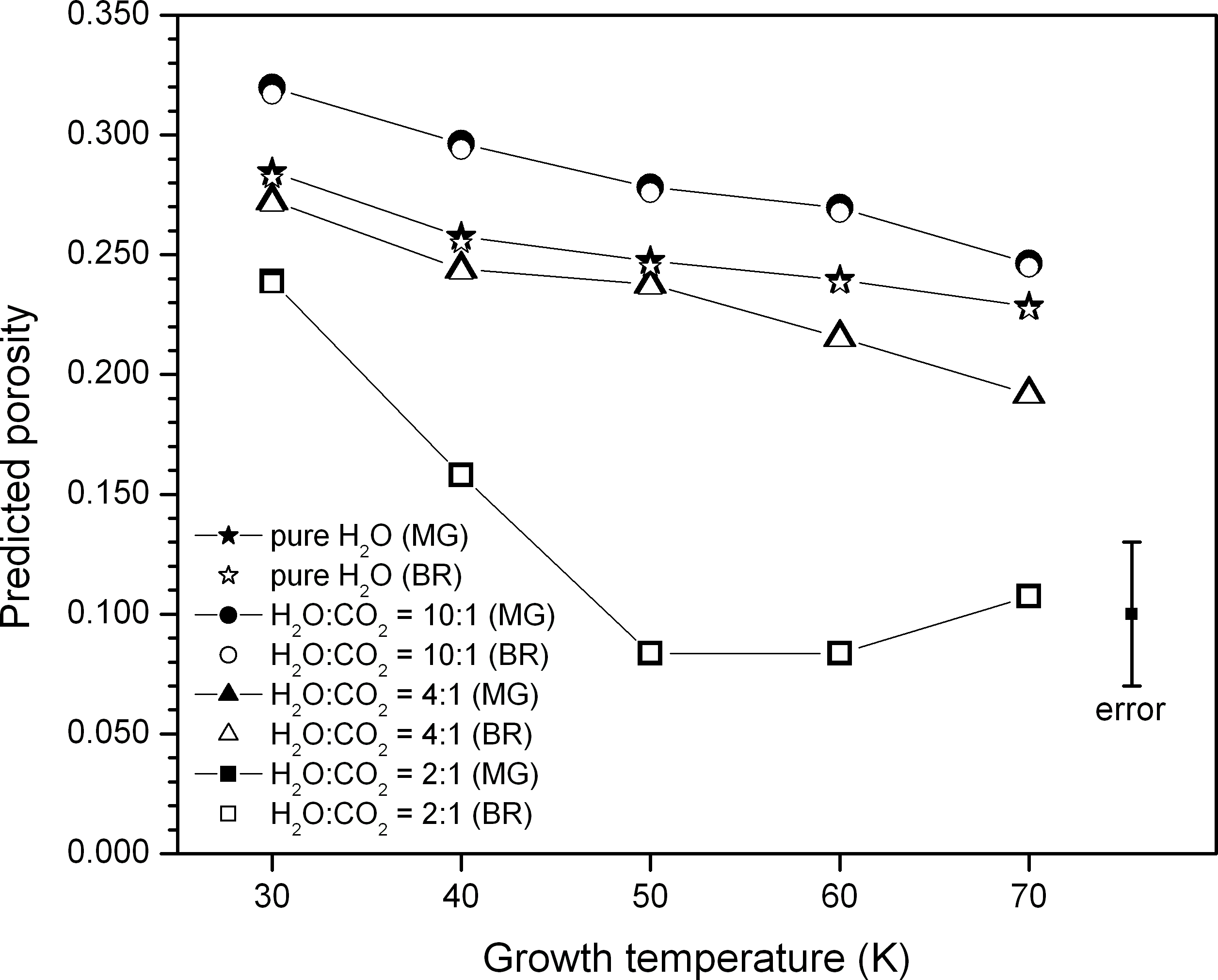} \caption{Predicted porosities of pure H$_{2}$O and H$_{2}$O:CO$_{2}$ ice samples as a function of the growth temperature, following the Maxwell-Garnett (MG) and the Bruggeman (BR) EMAs. Due to overlap the dark symbols cannot be distinguished well. The error in the prediced porosities (bottom right corner) depends on the error in the refractive indices (0.4\%) and on the error in the H$_{2}$O:CO$_{2}$ ratios (30\%).} \label{fig3}
\end{figure}

\indent Fig. \ref{fig2} shows the effective refractive indices predicted by the extended Maxwell-Garnett EMA as a function of the H$_{2}$O:CO$_{2}$ volume fraction ratio ($f_{C}/f_{B}$) and the porosity ($f_{A}$). We observe that the model is consistent with the boundary value conditions by reproducing the refractive indices for the nearly pure H$_2$O and CO$_2$ ice samples. The porosities can be determined from the measured refractive index and relative molecular abundances by finding the root that satisfies both $n_{eff}$(T) = $n_{1}$(T) and $f_{C}/f_{B}$ = 10, 4 or 2. Fig. \ref{fig3} shows the predicted porosities of pure H$_{2}$O (dark stars) ice samples obtained by the regular (i.e., non-extended) Maxwell-Garnett EMA, compared with the predicted porosities of H$_{2}$O:CO$_{2}$=10:1 (dark circles), H$_{2}$O:CO$_{2}$=4:1 (dark triangles), and H$_{2}$O:CO$_{2}$=2:1 (dark squares) ice samples as a function of the growth temperature. In general, we observe that the predicted porosities decrease when the growth temperature increases. The predicted porosities of the H$_2$O:CO$_2$=10:1 and 4:1 ice samples overlap within the experimental error and are rather close to the predicted porosities of the pure H$_2$O ice samples, suggesting that up to 20 \%, CO$_2$ does not significantly affect the ice morphology. In contrast, the predicted porosities involving the H$_2$O:CO$_2$=2:1 ice samples show a different behaviour when the growth temperature increases: at 40 K, the ice mixture seems to be slightly less porous than the pure H$_2$O ice sample grown within similar conditions. Beyond 40 K, the porosity decreases drastically until a plateau is reached from 50 K onwards, where the porosity loss can reach 65 \% relative to the pure H$_2$O ice sample. This noticeable gap suggests that the high abundance of CO$_2$ together with the growth temperature affect the ice morphology.                

\subsubsection{The extended Bruggeman EMA}
The Bruggeman dielectric function originally applies to a two component mixture \citep{Bohren83}. In this section, we use the second model proposed by \citep{Luo97} in which a three-component composite material is visualised as an aggregate structure where spheres of air (pores), spheres of bulk carbon dioxide ice, and spheres of bulk water ice are embedded in an effective medium, characterised by an effective dielectric constant ($\epsilon_{eff}$). In the following, we use the same $A$, $B$, $C$ nomenclature as in section \ref{extendedMG}. Hence for a porous H$_2$O:CO$_2$ ice sample of porosity $f_{A}$, deposited at a growth temperature T, the effective dielectric constant $\epsilon_{eff}$(T) can be determined by solving the following equation for different porosities ($0.001 \le f_{A} \le 0.998$) \citep{Luo97}
\begin{equation}
\label{eq19}
f_{A} \frac{\epsilon_{A}-\epsilon_{eff}}{\epsilon_{A}+2\epsilon_{eff}} + f_{B} \frac{\epsilon_{B}-\epsilon_{eff}}{\epsilon_{B}+2\epsilon_{eff}} + f_{C} \frac{\epsilon_{C}-\epsilon_{eff}}{\epsilon_{C}+2\epsilon_{eff}} = 0.
\end{equation}
As previously, we assume that the dielectric constant of the pores equals 1 and that the He-Ne light absorption in the ice sample is negligible, then Eq. \ref{eq19} becomes     
\begin{equation}
\label{eq20}
f_{A} \frac{1-n_{eff}^{2}}{1+2n_{eff}^{2}} + f_{B} \frac{n_{B}^{2}-n_{eff}^{2}}{n_{B}^{2}+2n_{eff}^{2}} + f_{C} \frac{n_{C}^{2}-n_{eff}^{2}}{n_{C}^{2}+2n_{eff}^{2}} = 0.
\end{equation}
Fig. \ref{fig3} shows the predicted porosities of H$_{2}$O:CO$_{2}$=10:1 (open circles), H$_{2}$O:CO$_{2}$=4:1 (open triangles), and H$_{2}$O:CO$_{2}$=2:1 (open squares) ice samples as a function of the growth temperature, compared with the predicted porosities of pure H$_{2}$O (open stars) ice samples obtained by the regular Bruggeman EMA. As seen previously for porous H$_{2}$O ice samples, the extended Bruggeman EMA predictions are similar to those of the extended Maxwell Garnett EMA. Therefore, the same conclusion as in the previous section with the extended Maxwell Garnett EMA can be drawn.  

\section{Discussion}
\label{section4}

\subsection{Benchmarking test: infrared spectroscopy}

\begin{figure}[]
\centering
\includegraphics[width=8.5cm]{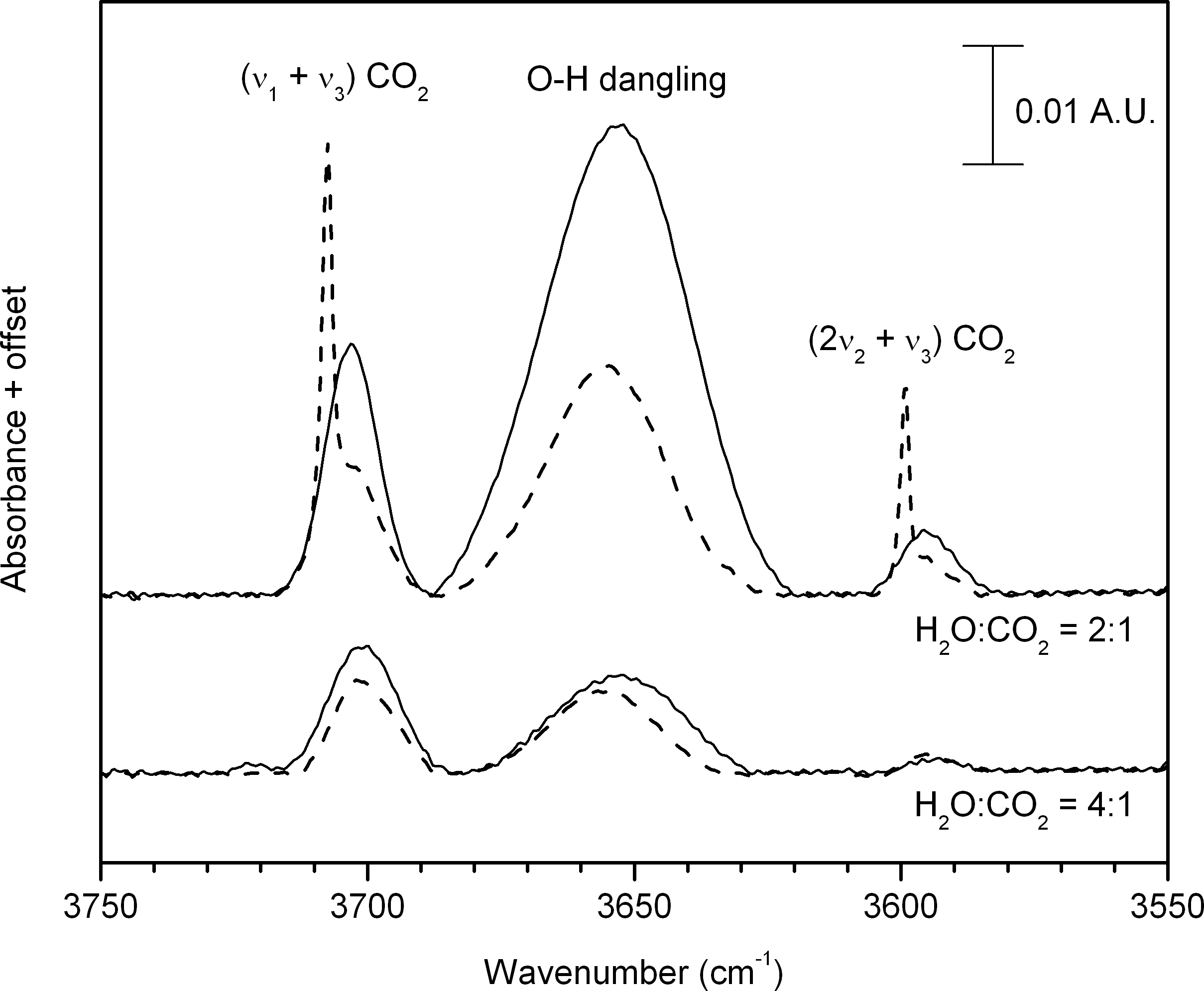} \caption{Infrared transmission spectra of H$_{2}$O:CO$_{2}$=4:1 and 2:1 ice samples grown by background deposition at 30 K (solid line) and 60 K (dash line). The absorption band related to the bulk O-H stretching modes has been subtracted for clarity.}
\label{fig5}
\end{figure}

\noindent The porosities predicted by the extended EMAs can be tested by using infrared spectroscopy. The infrared transmission spectra of H$_{2}$O:CO$_{2}$=4:1 and 2:1 ice samples grown by background deposition at 30 K (solid line) and 60 K (dash line) are depicted in Fig. \ref{fig5} in the 3750 -- 3550 cm$^{-1}$ range, covering the combination/overtone modes of carbon dioxide and the O-H dangling modes of water. The infrared spectra of H$_{2}$O:CO$_{2}$=10:1 ice samples are not shown here because of weak carbon dioxide absorption features. For the ice samples deposited at 30 K, two broad bands at 3702$\pm$1 cm$^{-1}$ ($\nu_{1}+\nu_{3}$) and 3594$\pm$1 cm$^{-1}$ (2$\nu_{2}+\nu_{3}$) are present and correspond to the combination/overtone modes of carbon dioxide \citep{Gerakines95,Bossa08}. For the H$_{2}$O:CO$_{2}$=2:1 ice sample deposited at 60 K, significant changes are observed: sharp features appear, overlapping the combination/overtone modes of carbon dioxide, which are attributed to pure CO$_{2}$ ice and indicative of segregation \citep{Hodyss08,Oberg09a}.\\
\indent Previous laboratory studies have reported that the band profile of the O-H dangling modes of the two- and three-coordinate surface-water molecules can shift or merge in one broad band, depending on the environment \citep{Rowland91,Palumbo06}. We observe a unique broad band at 3654$\pm$1 cm$^{-1}$ that we can attribute to the merged O-H dangling features of water \citep{Palumbo06}. This broad band provides information on the amount of pores in the ice \citep{Rowland91}. In order to qualitatively characterise the porosity difference between the ice samples, all spectra are corrected with a baseline, then normalised to the bulk O-H stretching modes (not shown here). Fig. \ref{fig5} shows that the intensity of the O-H dangling modes of a H$_{2}$O:CO$_{2}$=2:1 ice sample deposited at 60 K is significantly lower than the one of a same ice sample deposited at 30 K, meaning that a H$_{2}$O:CO$_{2}$=2:1 ice sample is less porous when deposited at 60 K compared to 30 K. In contrast, a H$_{2}$O:CO$_{2}$=4:1 ice sample does not show such large discrepancy in the intensity of the O-H dangling modes when deposited at 60 K compared to 30 K. Considering the integrated absorbance of the O-H dangling modes, the difference is about 60\% for the H$_{2}$O:CO$_{2}$=2:1 ice samples deposited 60 K compared to 30 K, and maximum 30\% for the H$_{2}$O:CO$_{2}$=4:1 ice samples deposited at 60 K compared to 30 K. Qualitatively, this is in good agreement with the predictions of the two extended EMAs. However, we cannot compare the two different ice mixtures one-to-one since the environment strongly affects the infrared band strengths and therefore the intensities.

\indent Palumbo 2006 has proposed that in H$_{2}$O:X ice mixtures (X = CO, CO$_2$ or CH$_4$), the X molecules can diffuse and stay in the pores, thus preventing a full compaction -- as seen for a pure H$_{2}$O ice sample -- after ion impact at low temperature. Here we observe after the background deposition that a compact structure appears together with segregation. We therefore propose a mechanism in which CO$_2$ molecules diffuse on the surface of the growing ice sample prior to be incorporated into the bulk, then fill the pores partly or completely, depending on the relative abundance of CO$_2$ and the growth temperature. In such a scheme, a large difference between the binding energies of the ice molecules (A-A, A-B, and B-B) will preferably form aggregates and the growth temperature will affect their diffusion. This is addressed in detail in a forthcoming paper by using Monte-Carlo simulations.

\subsection{Astrophysical implications}
The porosity of interstellar ices is a crucial parameter that defines the efficiency of the adsorption, the diffusion, the reaction, and the entrapment capacities of astrophysically relevant molecules. By providing large effective surface areas, pores have important consequences on the chemistry governed by surface processes. The presence of pores, as well as the evolution of the porosity during the different stages of star formations are key to understand the molecular complexity seen towards low mass and high mass protostars, but also in comets. Porous ices can undergo thermally induced structural collapse, thus affecting the diffusion of the interstellar ice components and therefore the catalytic properties \citep{Bossa12}. Previous investigations have shown that the morphology of a pure H$_2$O ice sample depends on the direction distribution of the incident gas phase molecules: omni-directional deposition leads to highly porous structures, in contrast to deposition along the surface normal that leads to more compact structures \citep{Kimmel01}. Omni-directional deposition is expected to occur in regions where non-thermal desorptions or exothermic solid state reactions eject icy material in the gas phase while keeping the grains temperature cold enough for the freeze-out of atoms and molecules. This type of deposition can occur during the formation of ices in a translucent cloud, but also in shock regions. Directional deposition is less likely but still possible when a particle or a body passes through a molecular cloud or a ring of a planet with a large relative velocity. A variety of other factors can influence the porosity, such as energetic environments that reduce pores or in the contrary enhance their formation like in cold environments. Morphology changes are expected when volatile species other than water are present during the omni-directional deposition as it is typically the case in the ISM.\\
\indent Since CO$_2$ ice is ubiquitous and abundant in the ISM, the study of ice mixtures produced by background deposition with different H$_{2}$O:CO$_{2}$ gas mixtures and growth temperatures is astronomically interesting. Our results demonstrate that segregation does affect the ice morphology. Since the segregation rates are temperature, thickness, and mixture dependent \citep{Oberg09a}, a wide range of porosities is expected depending on the environment. Premixed CO$_{2}$ with water in the gas phase does not significantly affect the ice morphology during omni-directional deposition as long as the physical conditions favourable to segregation are not reached, i.e., CO$_{2}$ abundances up to 20\% and a grain temperature below 30 K. Segregation can also occur with other volatiles species, such as CO and similar results for other systems like H$_{2}$O:CO or H$_{2}$O:CO:CO$_{2}$ ice mixtures are likely.\\
\indent Rare gas impurities premixed with water are expected to destabilise the hydrogen bonding network during the deposition at very low temperature (5 K) and then produce a highly porous structure \citep{Givan96}. In this study, we use CO$_{2}$ as the impurity and no porosity enhancement has been observed within the experimental error. However, surface-water molecules seem to be strongly affected by the presence of CO$_{2}$ since the O-H dangling modes are merged in an unique broad band, in contrast to a porous H$_{2}$O ice sample that shows two distinct absorption features  \citep{Rowland91}. Infrared spectroscopy of the O-H dangling modes has up to now been the unique tool for characterising the morphology of ices in space. However, their remote detections are made difficult by the weak intensities, the overlaps with the strong O-H stretching modes of water, and the band shapes that change depending on the environment. An alternative probing tool is therefore mandatory to assess the question of the porosity of inter- and circumstellar ices.\\
\indent Laboratory measurements of the complex refractive indices of astronomically relevant ices can be used to create model spectra in the mid and far infrared for comparison to spectra of interstellar and protostellar objects \citep{Hudgins1993}. In such studies, the optical constants can be used to simulate light scattering, light absorption, and light transmission, to calculate radiation transfer, and to determine the chemical composition of matter. Regarding this wide range of applications, the present results hold the potential to determine the porosity of inter- and circumstellar ices in a complementary way, by trying to match observational data and optical constants of multi-phase composite ices, using extended EMAs. We have shown that measuring the refractive index of a porous ice mixture in the visible also allows us to quantify its porosity. Refractive indices determination through visible polarimetry of icy materials onto the surface of satellites, comets \cite{Mukai87}, and interplanetary dust particles, therefore hold the potential to provide quantitative information on ice morphology following the methodology described in this study. It should be noted that for this, it is necessary that the relative abundances and intrinsic refractive indices of the different ice constituents are known.  




\section{Conclusions}
\label{section5}
We have measured the porosity of three astrophysically H$_{2}$O:CO$_{2}$ relevant ice mixtures grown by background deposition using a new laboratory-based method that combines optical laser interference and extended effective medium approximations. We find that: 
 
 \begin{enumerate} 
      \item Optical laser interference combined with extended effective medium approximations provide a tool for studying the porosity of ice mixtures grown by background deposition.
      \item Measuring the refractive index of a porous ice mixture in the visible also allows us to quantify its porosity if the relative abundances and intrinsic refractive indices of the different ice constituents are known.
      \item Premixed CO$_2$ with water in the gas phase does not significantly affect the ice morphology during omni-directional deposition as long as the physical conditions favourable to segregation are not reached.
      \item The three-phase layered model commonly used for refractive indices measurement is still valid for water-rich heterogeneous materials. 
   \end{enumerate}

\begin{acknowledgements}
Part of this work was supported by NOVA, the Netherlands Research School for Astronomy, a Vici grant from the Netherlands Organisation for Scientific Research (NWO), and the European Community 7$^{\tiny{th}}$ Framework Programme (FP7/2007-2013) under grant agreements no. 238258 (LASSIE) and no. 299258 (NATURALISM). We thank G. Strazzulla for providing the silicon substrate. We also thank S. Schlemmer for stimulating discussions and comments for this study. The referee is acknowledged for improving the contents of this paper.
\end{acknowledgements}

\clearpage



\end{document}